\documentclass{ws-ijmpe}
\usepackage[super,compress]{cite}
\usepackage{color}
\begin{document}

\markboth{Priyanka Sett, Prashant Shukla}
{Inferring freeze-out parameters from pion measurements at RHIC and LHC} 
%
\catchline{}{}{}{}{}
%

\title{Inferring freeze-out parameters from pion measurements at RHIC and LHC}

\author{Priyanka Sett}
\address{Nuclear Physics Division, Bhabha Atomic Research Center, Mumbai, 400085, India
sett.priyanka@gmail.com}

\author{Prashant Shukla}
\address{Nuclear Physics Division, Bhabha Atomic Research Center, Mumbai, 400085, India
pshuklabarc@gmail.com}

\maketitle

\begin{history}
\received{Day Month Year}
\revised{Day Month Year}
\end{history}

\begin{abstract}
 We analyze the transverse momentum spectra of charged pions measured in Au+Au collisions 
at $\sqrt{s_{NN}}$ = 200 GeV and in Pb+Pb collisions at $\sqrt{s_{NN}}$ = 2.76 TeV using the Tsallis 
distribution modified to include transverse flow. All the spectra are well described by the modified 
Tsallis distribution in an extended transverse momentum range upto 6 GeV/$c$. 
  The kinetic freeze-out temperature ($T$), average transverse flow ($\beta$) 
  and degree of non thermalization 
($q$) are obtained as a function of system size for both the energies. With increasing system size 
$\beta$ shows increasing trend whereas $T$ remains constant. 
  While the systems at RHIC and LHC energies show similar $\beta$ and $q$, the parameter $T$ is higher 
at LHC as compared to RHIC. 
 The kinetic freeze-out temperature is also extracted using the measured charged particle 
multiplicity and HBT volume of the system as a function of system size and collision energies. 

\end{abstract}

\ccode{PACS numbers:}

\section{Introduction}
\label{intro}
  The Quantum Chromodynamics (QCD), the theory of strong interaction suggests that at energy density above $\sim$ 1 GeV/fm$^3$ 
the hadronic matter undergoes a phase transition to Quark Gluon Plasma (QGP)~\cite{QGP}, a phase where the relevant  
degrees of freedom are quarks and gluons. 
  The heavy ion collisions at relativistic energy are the means to produce a large volume of hot/dense matter required to 
create and characterize such a phase~\cite{HIC,sqgp}.

 The quark gluon matter presumably with local thermal equilibrium expands hydrodynamically and undergoes a phase transition to 
hadronic matter which further cools till the multiple scatterings among particles are sufficient to keep 
them as one system. The hadrons then decouple from the system and their spectra would reflect the condition of the system 
at the time of freeze-out. 
  Hadrons (pions, kaons and protons) form the bulk of particles produced and are usually the first and easiest to be 
measured in a heavy ion collision experiment. 
  Traditionally, statistical model~\cite{stat_model} has been used at SPS and RHIC energies 
to infer the conditions at freeze-out using measured hadron ratios as input.
 Alternatively one can consider full transverse momentum ($p_T$) spectra of hadrons in heavy ion collisions. 
The bulk and collective effects~\cite{flow,recom} show up in the low and intermediate $p_T$ regions of hadron spectra 
while the high $p_T$ region above 5 GeV/$c$ consists of particles from jets which are produced in hard interactions.


  The Tsallis distribution~\cite{Tsallis, scale_ref} describes a system in terms of two 
parameters; temperature and $q$ which measures deviation from thermal distribution.
 It has been shown in Refs.~\cite{khandai_ijmpa, wong_wilk} that the 
functional form of the Tsallis distribution in terms of parameter $q$ is the same  
as the QCD-inspired Hagedorn formula~\cite{HAG1, HAGEFACT} in terms of power $n$.
  Both $n$ and $q$ are related and describe the power law tail of the 
hadron spectra coming from QCD hard scatterings. 
   The Tsallis distribution has been used extensively to describe the $p_T$ spectra of identified charged 
hadrons measured in $p+p$ collisions at RHIC and at LHC energies~\cite{khandai_ijmpa, ahep}.
 It does not always provide the best description of hadron $p_T$ distributions
in heavy ion collisions which are modified due to collective flow and thus Tsallis blast wave method is 
used as in Ref.~\cite{blastwave}. 
  The average transverse flow can be included in Tsallis distribution and keeping the functional 
form to be analytical as done in Refs.~\cite{khandai_jpg, bhaskarde_epj}. 
  The function presented in Ref.~\cite{khandai_jpg}  can be used in a wider $p_T$ range as was done for 
both meson and baryon spectra for Au+Au collisions at $\sqrt{s_{NN}}$ = 200 GeV.

 We analyze the transverse momentum spectra of charged pions measured in heavy ion collisions. 
 Recent measurements of identified charged particle spectra by PHENIX in different centralities 
of Au+Au collision at $\sqrt{s_{NN}}$ = 200 GeV~\cite{PPG146} and by ALICE in  
the most central (0-5\%) and the most peripheral (60-80\%) Pb+Pb collisions 
at $\sqrt{s_{NN}}$ = 2.76 TeV~\cite{alice_chargedhadrons_2.76PbPb_highpT} 
have been used in the study.
  The kinetic freeze-out temperature ($T$), average transverse flow ($\beta$) 
  and degree of non thermalization 
($q$) are obtained as a function of system size for both the energies. 
  As an alternative the (kinetic) freeze-out temperature is also extracted using the measured charged 
particle multiplicity and HBT volume of the system. 

\section{Analysis of hadron spectra}
\label{formulation}
 The transverse momentum spectra of hadrons can be described using the 
modified Tsallis distribution including the transverse flow as  proposed in Ref.~\cite{khandai_jpg} is given by : 

\begin{eqnarray}
E\frac{d^3N}{dp^3}  =  C_{n}\left({\rm exp}\left(\frac {-\gamma \beta p_{T}} {nT}\right) + \frac {\gamma m_{T}}{nT}\right)^{-n}.
\label{Tsallis_mod}
\end{eqnarray}
Here $C_{n}$ is the normalization constant, $m_T=\sqrt{p_T^2 + m^2}$, 
$\gamma = 1/\sqrt{1-\beta^{2}}$, $\beta$ is the average transverse velocity of the system and 
$T$ is the temperature. 
The power $n$ is related to the non-extensivity parameter $q$ as $n = 1/(q - 1)$. 
The parameter $q$ gives temperature fluctuations~\cite{q_Tsallis} in the system as: $q-1 = Var(T)/\langle T \rangle^{2}$. 
It can take a value between 1 and 4/3.
   Larger values of $q$ correspond to smaller values of $n$ which imply dominant hard 
QCD point-like scattering. Both $n$ and $q$ have been interchangeably used in 
Tsallis distribution~\cite{scale_ref, PPG099, Phenix, Star, cleymans}. In heavy ion collisions, 
the high $p_T$ tail decides the value of $n$. 
 Phenomenological studies suggest that, for quark-quark point scattering, $n\sim$ 4~\cite{BlankenbeclerPRD12, BrodskyPLB637}, 
and when multiple scattering centers are involved $n$ grows larger.
When $\beta$ is zero, Eq.~\ref{Tsallis_mod} is the usual Tsallis equation which
has been the most popular tool to characterize hadronic collisions~\cite{wong_wilk, q_Tsallis, cleymans, cleymans2} in 
recent years.
  At low $p_T$, Eq.~\ref{Tsallis_mod} represents a thermalized system with collective flow and at high $p_T$
it becomes a power law as follows 
\begin{eqnarray}
\label{boltz}
 E\frac{d^3N}{dp^3}
  &\simeq & C_n \exp\left(\frac {-\gamma(m_{T} - \beta p_T)} {T}\right)
 \,\,\,\,{\rm for}\,\,\,\,\, p_{T} \rightarrow 0  \nonumber, \\
  &\simeq & C_n \left(\frac  {\gamma m_{T}} {nT} \right)^{-n}  \,\,\,{\rm for}\,\,\, p_{T} \rightarrow \infty.
\label{power}
\end{eqnarray}  

 In this work, we focus on the study of the charged pion spectra measured in heavy-ion collisions 
at RHIC and LHC energies. The errors on the data are taken as quadratic sums 
of statistical and uncorrelated systematic errors. The RHIC measurements are available in 
$p_T$ range 0.5 - 6.0 GeV/$c$ and we use the LHC measurements in the same range.
 The spectra are fitted with Eq.~\ref{Tsallis_mod} and all the parameters are obtained as a 
function of system size (centrality) for both the energies. 

\begin{figure}
\begin{center}
  \includegraphics[width=0.6\textwidth]{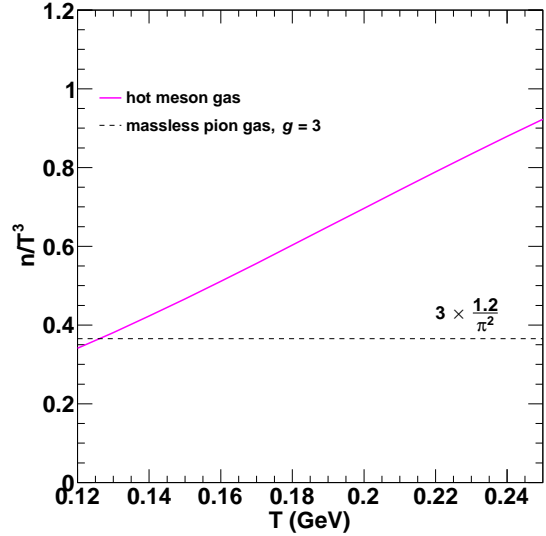} 
\caption{(Color online) The variation of $n/T^3$ as a function of 
temperature for both hot meson gas (solid line) and massless pion gas (dashed line).}
\label{fig:f_ah_noden} 
\end{center} 
\end{figure}

  The freeze-out temperature $T$ can also be extracted from the measured 
multiplicity using following procedure. 
 The particle number density $n$ can be related to the measured particle multiplicity and 
HBT volume $V$ as 
\begin{eqnarray}
\label{e1}
n\, V = \frac{dN}{d\eta}, 
\end{eqnarray}
where $dN/d\eta$ is 1.5 times the total measured 
charged particle multiplicty ($dN_{ch}/d\eta$). 
The number density can also be expressed in  terms of freeze-out temperature $T$  
\begin{eqnarray}
\label{e2}
n = \frac{1.2}{\pi^2} \, a_n (T) \,T^3.
\end{eqnarray}
 The parameter $a_n\,(T)$  = $\sum_i \, g_i n_i(m_i/T)$ where $g_i$ is the degeneracy factor 
and $n_i$ for $i^{\rm th}$ meson species is given by   
\begin{eqnarray}
\label{e9}
n_i (m_i/T) = \frac{1}{2\,\times\,1.2} \int_0^{\infty} \frac{x^2\, dx}{e^{(\sqrt{x^2+(m_i/T)^2}} - 1}. 
\end{eqnarray}
 The parameter $a_n\,(T)$  = 3 for massless pion gas. 
In our study we assume that the system at freeze-out consists of  
pion ($g$ = 3), kaon ($g$ = 4), $\rho$ ($g$ = 9), $\phi$ ($g$ = 1), 
$\eta$ ($g$ = 1), $\omega$ ($g$ = 3) mesons 
and obtained $n/T^3$ as a function of temperature which 
is shown in Fig.~\ref{fig:f_ah_noden} along with that for massless pion gas.

The freeze-out temperature $T$ can be obtained by numerically solving the following
equation 
\begin{eqnarray}
\label{e10}
T^3 = \frac{1}{(1.2/\pi^2) \, a_n(T)} \, \frac{1}{V}\, \frac{dN}{d\eta}. 
\end{eqnarray}
The HBT volume $V = (2\pi)^{3/2}$ $R_{side}^2\, R_{long}$, 
where $R_{side}$ and  $R_{long}$ are the 
measured HBT radii~\cite{star_twopion_interferometry,  alice_twopion_interferometry}.


 If transverse flow is present in the system, then the 
system volume obtained from measured HBT radii will be smaller than the 
fireball volume. To correct for this effect the HBT radii as a function of 
$m_T$ are extrapolated to $m_T$ = 0.140 GeV/$c^2$ which corresponds to 
$p_T$ = 0.

\begin{figure}
\begin{center}
  \includegraphics[width=0.6\textwidth]{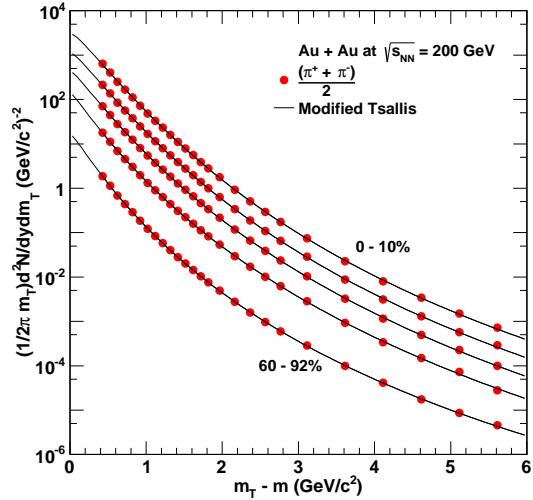}
  \caption{(Color online) Invariant yield for charged pions~\cite{PPG146}
    as a function of ($m_T$ - $m$)  measured in Au+Au collisions at 
    $\sqrt{s_{NN}}$ = 200 GeV for 0-10\%, 10-20\%, 20-40\%, 40-60\%  and 60-92\% centrality bins. 
    The spectra are scaled up by a factor of 10, 5, 3, 2 and 1 for the respective
    centrality bins. The fitted Modified Tsallis function is shown by the black curve.}
  \label{fig:pi_200_auau} 
\end{center} 
\end{figure}
\begin{figure}
\begin{center}
\includegraphics[width=0.6\textwidth]{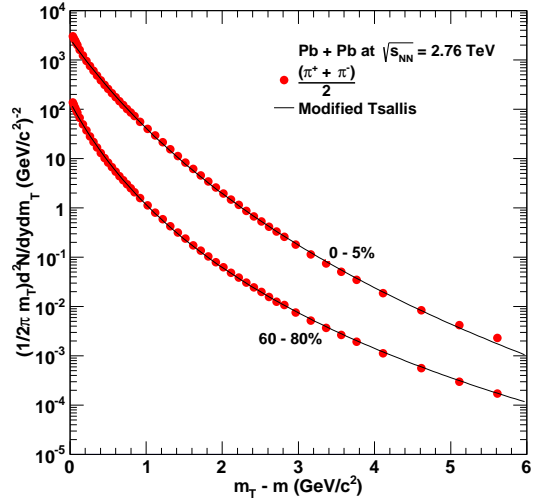}
\caption{(Color online) Invariant yield for charged pions~\cite{alice_chargedhadrons_2.76PbPb_highpT} 
  as a function of ($m_T$ - $m$) measured in Pb+Pb collisions at $\sqrt{s_{NN}}$ = 2.76 TeV 
  for 0-5\% and 60-80\% centrality bins. The Modified Tsallis fitted function is shown by the black curve.}
\label{fig:pi_2.76_pbpb} 
\end{center} 
\end{figure}

\begin{figure*}
\begin{center}
\includegraphics[width=0.49\textwidth]{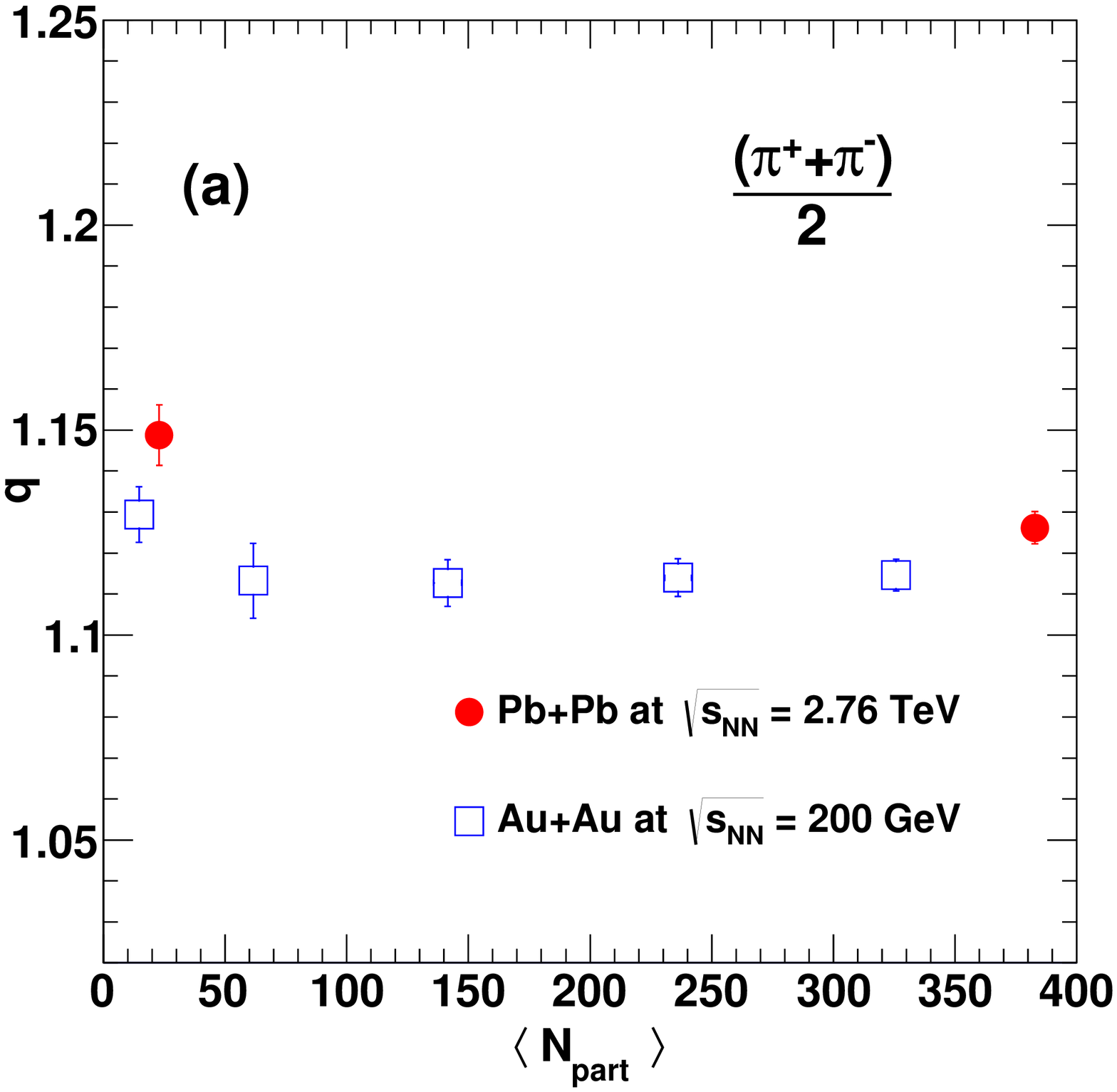}
\includegraphics[width=0.49\textwidth]{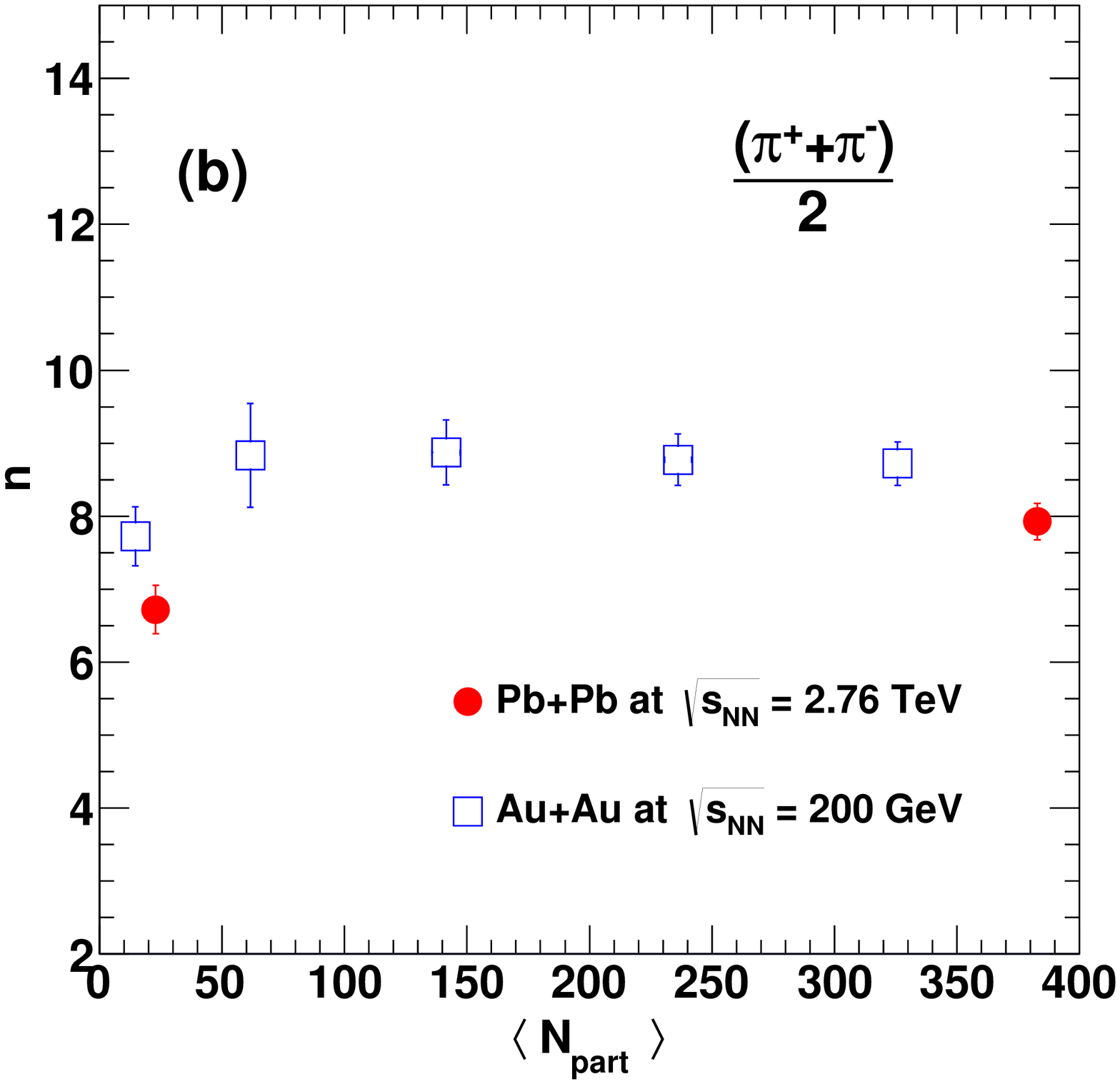}
\caption{(Color online) Parameters (a) $q$ and (b) $n$ obtained from Eq.~\ref{Tsallis_mod} for charged pions 
  measured in Au+Au (open squares) collisions at $\sqrt{s_{NN}}$ = 200 GeV and in Pb+Pb (filled circles) 
  collisions at $\sqrt{s_{NN}}$ = 2.76 TeV as a function of  $\langle N_{part} \rangle$. }
\label{fig:n_q} 
\end{center} 
\end{figure*}

\section{Results and Discussions}
\label{results}


\begin{figure*}
\begin{center}
\includegraphics[width=0.49\textwidth]{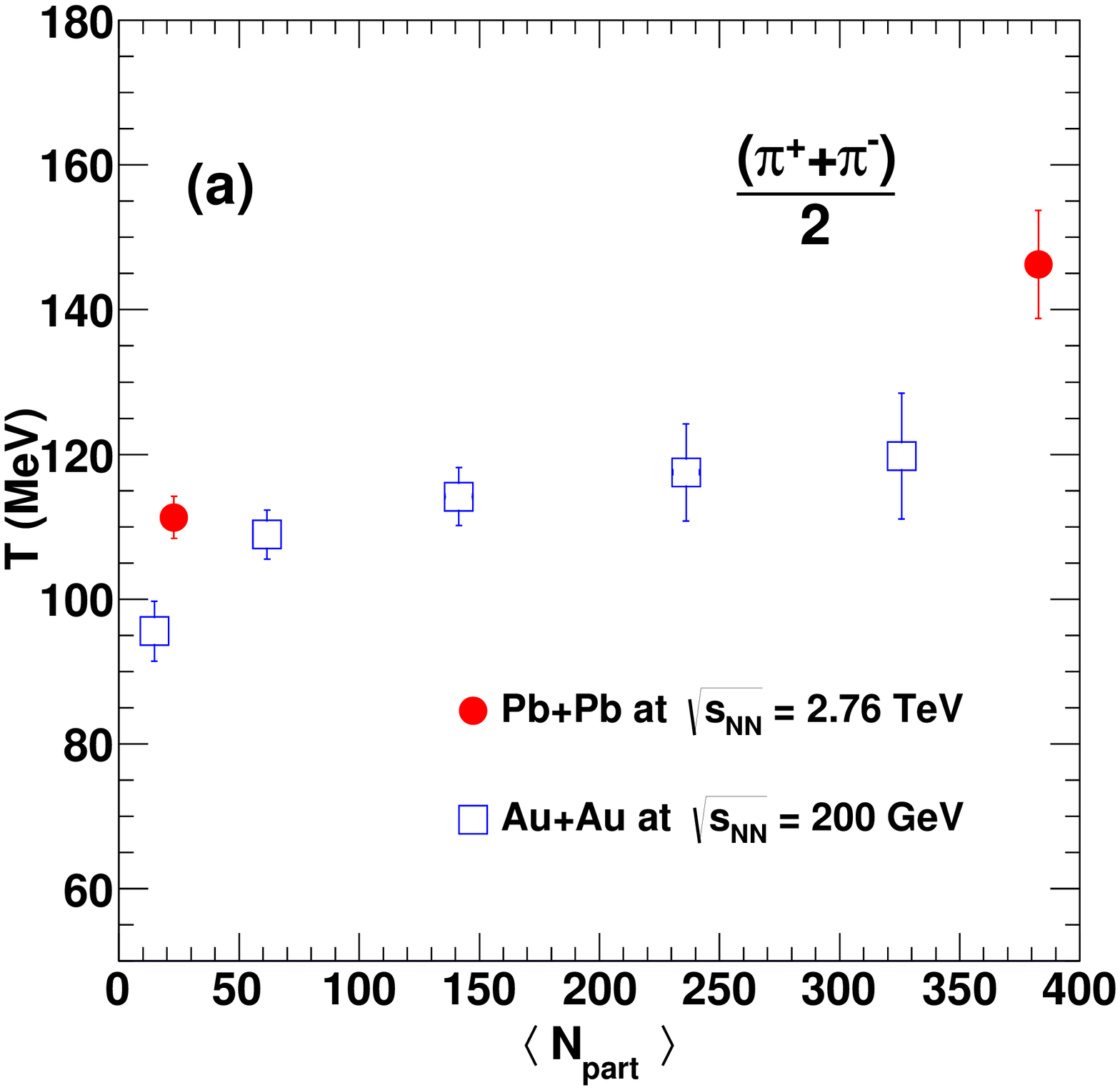}
\includegraphics[width=0.49\textwidth]{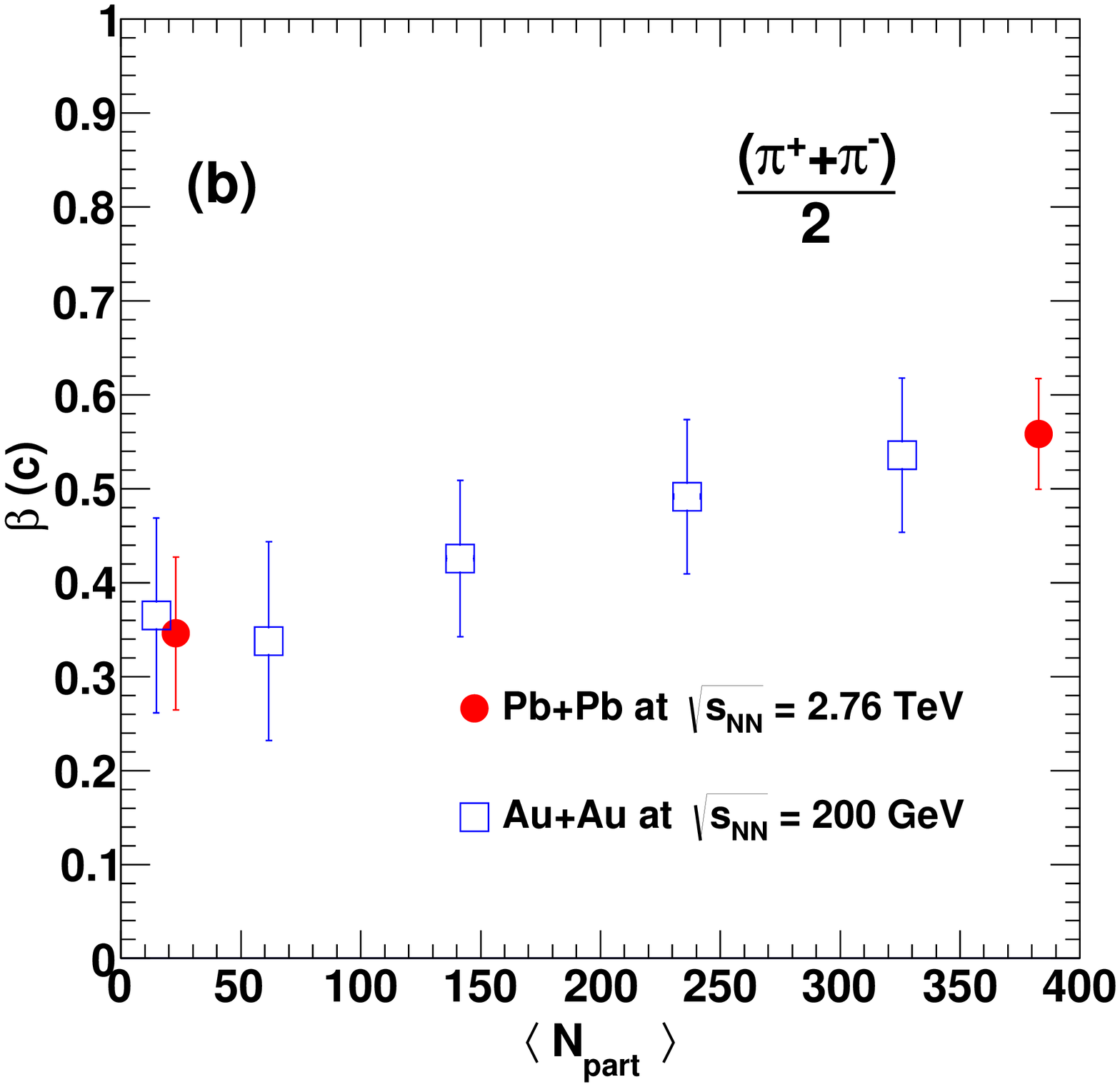}
\caption{(Color online) Parameters (a) $T$ and (b) $\beta$ obtained from Eq.~\ref{Tsallis_mod} for charged pions 
  measured in  Au+Au (open squares) collisions at $\sqrt{s_{NN}}$ = 200 GeV and Pb+Pb (filled circles) 
  collisions at $\sqrt{s_{NN}}$ = 2.76 TeV as a function of  $\langle N_{part} \rangle$. }
\label{fig:T_beta} 
\end{center} 
\end{figure*}

\begin{figure*}
\begin{center}
\includegraphics[width=0.49\textwidth]{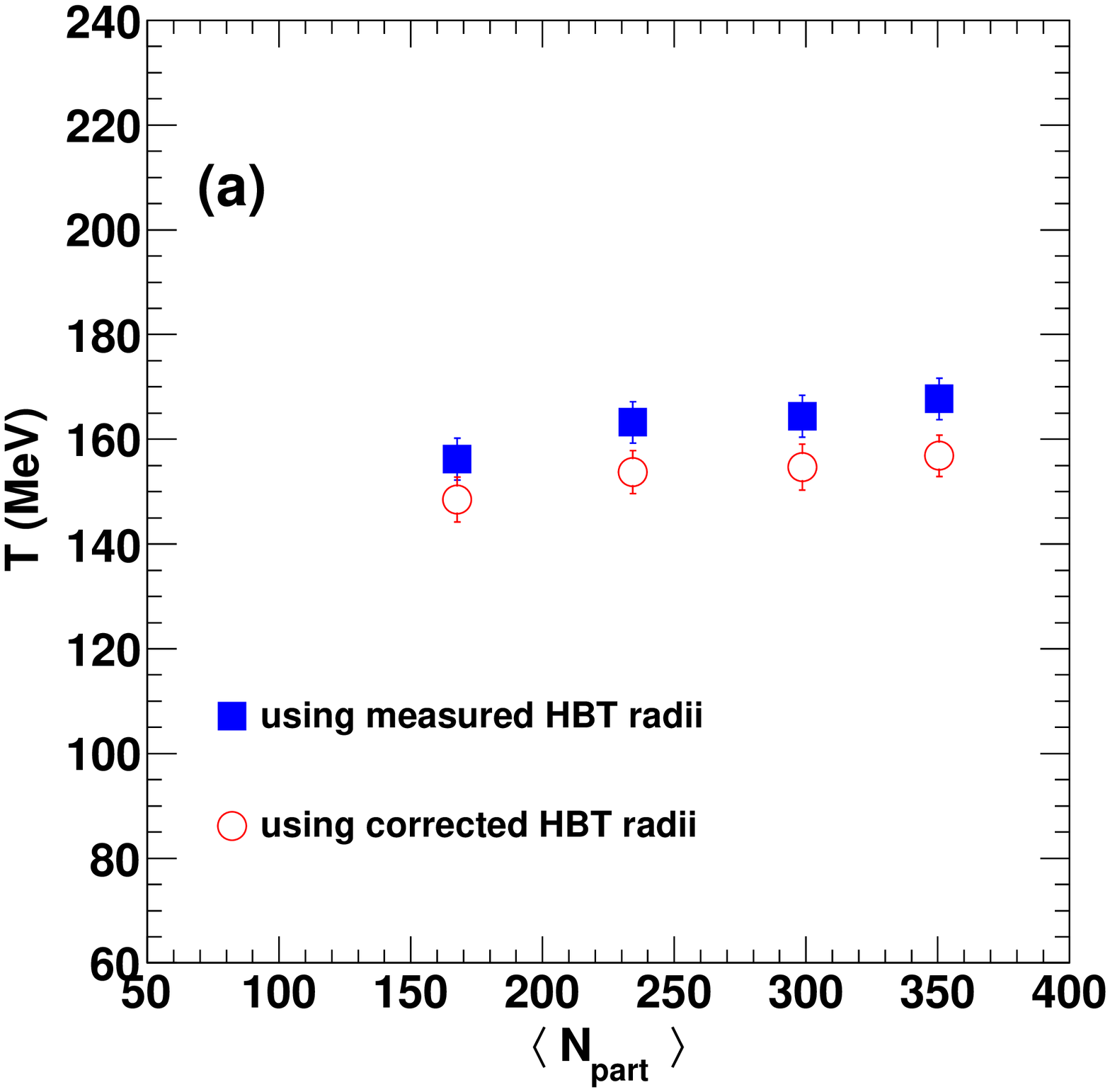}
\includegraphics[width=0.49\textwidth]{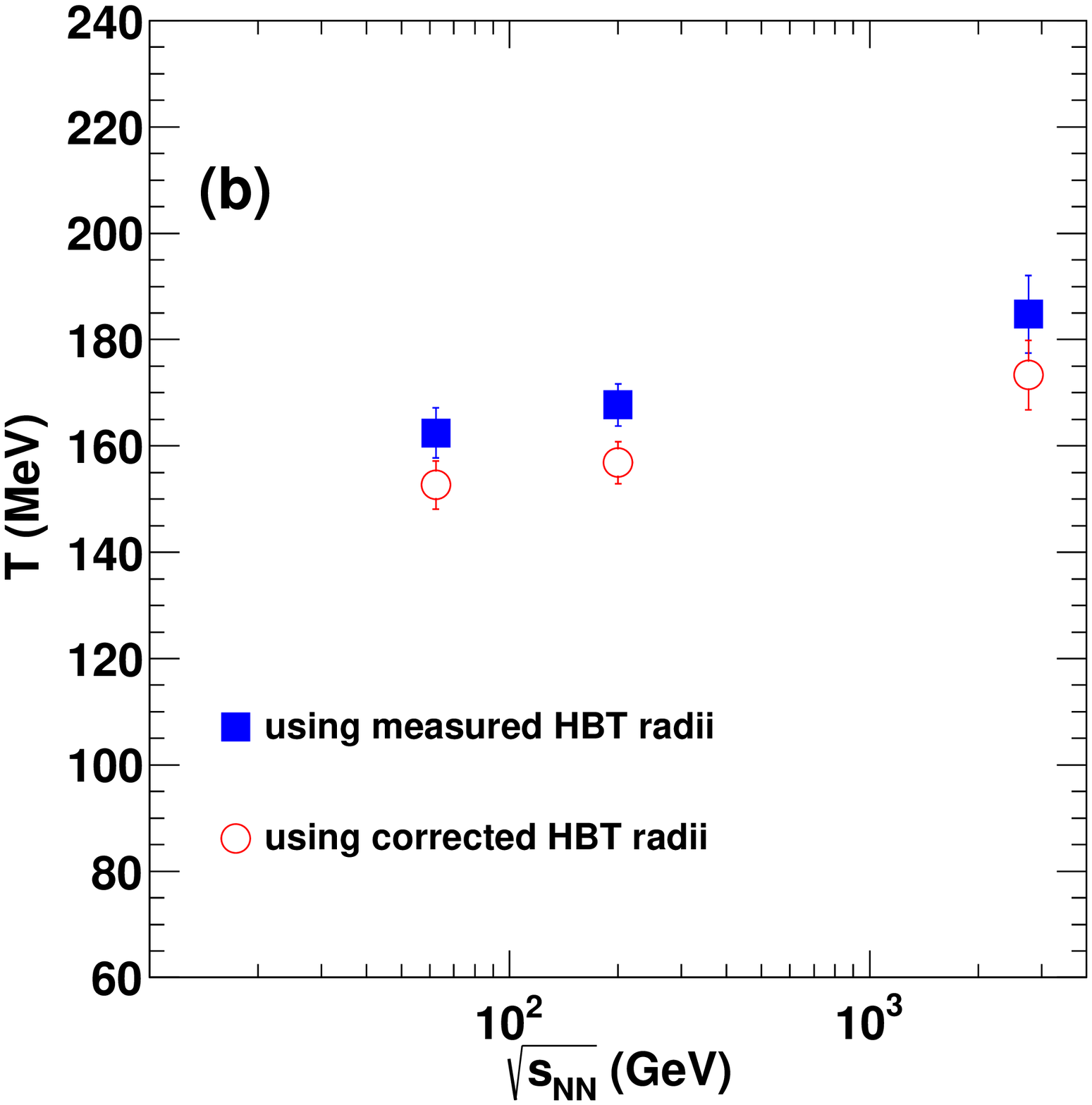}
\caption{(Color online) Kinetic freeze-out temperature obtained using Eq.~\ref{e10} (a) as a function of 
$\langle N_{part} \rangle$ and (b) as a function of $\sqrt{s_{NN}}$ (0-5\% centrality) for RHIC and LHC energies. }
\label{fig:T_npart_snn} 
\end{center} 
\end{figure*}



   Figure~\ref{fig:pi_200_auau} shows the charged pion invariant yield spectra
in Au+Au collisions at $\sqrt{s_{NN}}$ = 200 GeV~\cite{PPG146} as a function of ($m_T$ - $m$), fitted 
with Eq.~\ref{Tsallis_mod}. The spectra in Au+Au collisions are given for 0-10\%, 
10-20\%, 20-40\%, 40-60\% and 60-80\% centralities, which are scaled up by factors given in the 
Fig.~\ref{fig:pi_200_auau}. 
Figure~\ref{fig:pi_2.76_pbpb} shows the same for Pb+Pb collisions at 
$\sqrt{s_{NN}}$ = 2.76~\cite{alice_chargedhadrons_2.76PbPb_highpT}  
TeV fitted with Eq.~\ref{Tsallis_mod}. 
It is seen that Eq.~\ref{Tsallis_mod} describes the data in full $p_T$ range for all 
collision centralities, both at RHIC and LHC energies. 
  The collision centralities can be converted to average number of participants $\langle N_{part} \rangle$
using Glauber model which is proportional to initial system size.

 The parameter $q$ as a function of $\langle N_{part} \rangle$ is shown in Fig.~\ref{fig:n_q} (a). 
The value of $q$ is higher for the peripheral collisions in comparision to other centralities, 
for both RHIC and LHC energies. The corresponding value for power $n$ is shown in 
Fig.~\ref{fig:n_q} (b).
  The values of $q$ (or $n$) are similar for RHIC and LHC which show similar degrees of 
thermalization at the two energy regimes.

 Figure~\ref{fig:T_beta} (a) and \ref{fig:T_beta} (b) respectively show the kinetic freeze-out 
temperature ($T$) and the average transverse flow ($\beta$) as a function of $\langle N_{part} \rangle$.  
At RHIC energy, except for peripheral bin, the value of temperature remains constant 
for all centrality bins within uncertainties. 
For the most central collisions (0-5\%), $T$ has a higher value at LHC energy than that  at RHIC. 
  The transverse flow velocity $\beta$ increases with the system size for both RHIC and LHC collisions
and has almost same behavior in the two energy regimes.


The freeze-out temperature is obtained from Eq.~\ref{e10} using the 
measured particle multiplicity and HBT volume.
Figure~\ref{fig:T_npart_snn} (a) shows the freeze-out temperature as a function of 
system size and Fig.~\ref{fig:T_npart_snn} (b) shows the same as a function of 
collision energy. The solid squares show the result obtained from the measured 
HBT radii. The open circles show the result obtained using the corrected HBT radii 
(at $p_T$ = 0) as explained before. This correction makes the 
values of $T$ slightly smaller. It is seen that the freeze-out temperature 
increases while going from RHIC energy to LHC energy. 

A comparison of Figs.~\ref{fig:T_beta} (a) and \ref{fig:T_npart_snn} (a) 
shows that the freeze-out temperature obtained from the two different methods 
follow similar trend as a function of system size. The temperature shows almost flat behavior 
with system size. 
Figures~\ref{fig:T_beta} (a) and \ref{fig:T_npart_snn} (b) show that  
the freeze-out temperature is more for LHC energy.  
 The temperature obtained from two measurements namely the $p_T$ spectra and the 
HBT measurements have same dependence on system size and energy. 
 There is upto 20\% difference in the magnitudes which might arise due to the
experimental error and different $p_T$ range of the measurements affected by 
transverse flow differently.



\section{Conclusion}

   The transverse momentum spectra of charged pions measured in Au+Au collisions 
at $\sqrt{s_{NN}}$ = 200 GeV and Pb+Pb collisions at $\sqrt{s_{NN}}$ = 2.76 TeV are analysed 
using the modified Tsallis distribution. 
 All the spectra used in this analysis are well described by this distribution. 
  The parameter $q$ of the modified Tsallis distribution suggests similar 
thermalization characteristics for systems at RHIC and LHC energies.
 The kinetic freeze-out temperature extracted from pion $p_T$ spectra remains flat for all 
centrality bins except for the peripheral bin at RHIC energy. 
The freeze-out temperature is higher at LHC energy than that at RHIC energy for the 
most central collisions.  
  The transverse flow velocity increases with system size for both of these energies. 
The kinetic freeze-out temperature is obtained as a function of 
system size and collision energy, from the measurement of HBT radii 
and particle multiplicity. The measured HBT radii 
are extrapolated to $m_T=0.140$ GeV/$c^2$ to correct for the effect of 
transverse flow. 
The freeze-out temperature obtained from particle spectra and HBT measurements show similar trend as 
a function of system size as well as a function of collision energies. However, 
the kinetic freeze-out temperature obtained using HBT radii 
remains larger than that obtained from the particle spectra.



\end{document}